\documentstyle[11pt] {article}

\title{Signed Phases and Fields Associated with Degeneracies}

\author{R. Englman$^{a,b}$, A. Yahalom$^a$ \\
$^a$ College of Judea and Samaria, Ariel 44284, Israel\\
$^b$ Department of Physics and Applied Mathematics,\\
Soreq NRC,Yavne 81800,Israel\\
e-mail: englman@vms.huji.ac.il; asya@ycariel.yosh.ac.il;}

\begin{document}

\maketitle

\newcommand{\beq} {\begin{equation}}
\newcommand{\enq} {\end{equation}}
\newcommand{\ber} {\begin {eqnarray}}
\newcommand{\enr} {\end {eqnarray}}
\newcommand{\eq} {equation}
\newcommand{\eqs} {equations }
\newcommand{\mn}  {{\mu \nu}}
\newcommand{\sn}  {{\sigma \nu}}
\newcommand{\rhm}  {{\rho \mu}}
\newcommand{\sr}  {{\sigma \rho}}
\newcommand{\bh}  {{\bar h}}
\newcommand {\er}[1] {equation (\ref{#1}) }
\newcommand {\SE} {Schr\"{o}dinger equation}

\begin {abstract}
In the first part, expressions are given for the {\it sign} of the
topological angle that is acquired
 upon making a loop around a degeneracy ("conical intersection") point
of two molecular energy surfaces. The expressions involve the
 partial derivatives (with respect to the nuclear coordinates) of the
matrix
 elements of the coupling Hamiltonian. Examples are given of a few
studied
cases, such as of excited states  that have topological angles with a
sign opposite to those in the
 ground states.

In the second part, the two dimensional (or two parameter) situation
that
 characterizes a conical intersection (ci) between potential surfaces
in a
 polyatomic molecule is constructed as a limiting case of the three
 dimensional Dirac-monopole situation. For an electron occupying a
twofold
 state, we obtain both the "magnetic-field" (or curl-field) and the
tensorial
(or Yang-Mills-) field (which  is the sum of a curl and of a vector-
product
term). These pseudo- fields represent the reaction  of the electron on
the
 nuclear motion via the nonadiabatic coupling terms (NACTs). We find
that both
 fields are aligned with the orthogonal, (so called) seam directions of
the
 ci and are zero everywhere outside the seam, but they differ as
regards
 the flux that they produce. In a two-state situation, the fields
 are representation dependent and the values of, e.g., the fluxes
depend on
 the state that the electron occupies. The angular dependence
of the NACTs and the fields calculated from a general linearly coupled
model
 agrees with recently computed results for $C_2 H$ [A.M. Mebel, M. Baer
  and S.H. Lin, J.Chem. Phys. {\bf 115} 3673 (2001)].
An effective-Hamiltonian formalism is proposed for experimentally
observing and
distinguishing between the different fields.
\end {abstract}
\section {Introduction}
In 1975 Longuet-Higgins showed that if there is a sign change in the
wave
function upon  performing a closed loop in the parameter space, then a
point of
 degeneracy is enclosed by the loop \cite {LonguetH}. Trivially, a sign
change amounts to
a phase angle change of ${\pm} (2N+1)\pi $, where $N$ is an integer or
zero  and is, in fact, zero when the circulation is about a
degeneracy point which is a single conical intersection ({\it ci}). In
the following section of this work
 we consider the sign of the added phase angle change, first in a
general
 manner, then for a special ("the complex") representation, including
several examples that have been discussed in the literature (\cite
{ZwanzigerG} -
\cite {EYACP}). The sign associated with any  single {\it ci} will be
of added
 interest in cases that there are several {\it ci}'s in the system. We
 shall focus attention on such multiple {\it ci} situations.
The next section has been motivated by some recent publications that
connect the electron-nucleus interaction with the three-dimensional
Dirac-monopole formalism \cite {MatsikaY} and with a pseudo-magnetic
field \cite {Baer2001} due to the nonadiabatic coupling. In the latter
paper
 an angular dependence of the field was postulated. We now  {\it
derive}
 and specify the unique form of this field for the most general model
 having a linear form of electron-nuclear interaction near the conical
intersection.
A further pseudo-field (the Yang-Mills tensorial field) is also introduced in the
present molecular
context and is evaluated.
\section {Signs of Geometrical Phases around Conical Intersections}

\subsection {Cartesian, real representation}

Let $X$ and $Y$ denote the coordinates of the parameter-plane in which
the
circling is being performed and let ($X_0 ,Y_0$) denote the location of
a
 {\it ci}. We shall
consider  the case where the potential surfaces of only two states
of the system cross at
this {\it ci}. In the vicinity of the {\it ci} we are entitled to
disregard possible
 interactions with all  other states and to write
the potential energy $ V(X,Y) $ as a $2 \times 2$ matrix that has
 the following form.

\beq
V(X,Y)= \left(\begin{array}{cc}
           -A(X,Y) & B(X,Y)\\
           B(X,Y) & A(X,Y)
\end{array} \right)\qquad
\label {potential}
\enq

In the absence of a magnetic field the matrix components can be taken
to be real.
A scalar term has been omitted, since this would not affect our
considerations.
At a {\it ci} we have
\beq
A(X_0 ,Y_0) = 0,~  B(X_0 ,Y_0) = 0
\label {AB}
\enq
but the partial derivatives of either $A$ or $B$ with respect
 to the coordinates (to be designated by subscripts $X$ and $Y$) cannot
both be zero
at $ (X_0 ,Y_0)$. (Otherwise, the intersection is touching, not
conical.)
The  mixing angle in the eigenstates of the potential matrix
 \er{potential} is given by
\ber
\theta & = & {1\over 2} arctan \frac{B(X,Y)}{A(X,Y)}
\nonumber \\
       & \approx & {1\over 2} arctan \frac{(X-X_0 )B_X (X_0 ,Y_0) + (Y-
Y_0 )B_Y (X_0 ,Y_0)}
{(X-X_0 )A_X (X_0 ,Y_0) + (Y-Y_0 )A_Y (X_0 ,Y_0)}
\label{theta1}
\enr
the second line being valid close to the intersection point. Use has
been made of \er {AB}. We now
 introduce
 the radius of circling $\Delta $ and the circling angle $\alpha $
shown
 in the figure, for the case that $A_X$ is non-zero.

\begin{figure}
\vspace{4cm}
\begin{picture}(1,1)
\end{picture}
\includegraphics{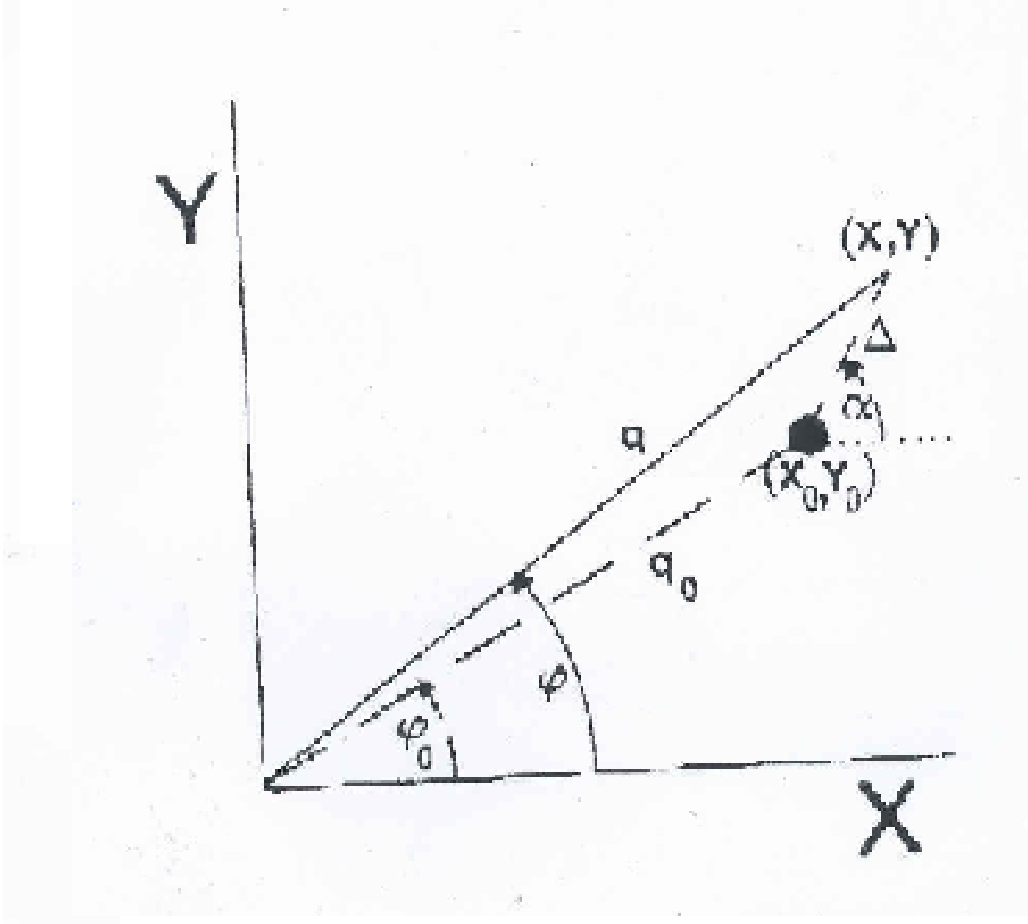}
\caption {Coordinate systems for the point $ (X,Y) $ or $ (q,\phi ) $
that circles around
the point of conical intersection [the dot located at $ (X_0,Y_0) $ or
$
(q_0,\phi_0 )$ ]. The circling is with a radius $ \Delta $ and
circling angle $ \alpha $.}
\label {fig1}
\end{figure}

 (If $A_X=0$, then
$A_Y$ must be non zero and we use an angle oriented by ${1\over 2} \pi
$ from
that shown in the drawing.) The introduced quantities are implicitly
defined
by
\beq
X-X_0  = \Delta \cos \alpha,~
Y-Y_0  = \Delta \sin \alpha
\label {deltaalpha}
\enq
 We now note the important fact that for a {\it ci}, in its vicinity,
the acquired phase
 is monotonic with the circling angle and  it is therefore permissible
to
obtain the sign of the phase by looking merely at the  limits
\beq
 \alpha \to 0
\label {limit1}
\enq
or, equivalently,
\beq
\cos \alpha \to 1,~~  \sin \alpha \to \alpha
\label {limit2}
\enq
We thus find for the phase angle, after some elementary simplification,
\beq
\theta \approx  {1\over 2 } \arctan (\frac{B_X}{A_X} + \alpha \frac
{A_X B_Y
-B_X A_Y}{A_X^2})
\label {thetaapprox}
\enq
with all derivatives to be evaluated at the {\it ci}. Since the inverse
tangent is an
increasing function of it argument, as the circling angle $\alpha $
increases,
the mixing angle $\theta $ will increase or decrease  depending on
whether
\beq
A_X B_Y-B_X A_Y > or < 0
\label {xineq}
\enq
(We recall that $A_X$ was assumed to be non-zero).

Example 1:

A pair of conical intersections, e.g.,
(a) in some bent molecules, as $H_2 S$ of $C_{2v} $ symmetry, two {\it
ci}'s
 between the lowest excited states $^1 A_2$ and $^1 B_2$,
(b)   for $C_2 H$ in $C_s $ symmetry
 the $3 ^2 A'$ and $4 ^2 A'$ states  \cite {MebelBL2000,MebelBL2001}

We now write out the Hamiltonian representing the coupling between
the nuclei, whose coordinates are denoted by $(X,Y)$ and the two
electronic
states whose surfaces are intersecting conically. For the case of a
pair of
{\it ci} symmetrically situated in the $X,Y$-plane the situation can be
 described by the following matrix elements:
\ber
A(X,Y) = (X^2 -1),~
B(X,Y) = Y
\label {AB1}
\enr

We have so shifted and scaled the nuclear coordinates, that the {\it
ci}
 are located at $X=\pm 1, Y=0$. The representation symbols of Herzberg
\cite {Herzberg} have been used, for which $X$ transforms in $C_{2v}$
as
$B_2$, $Y$ as  $B_1$ and $Z$ as $A_1$. We verify that at both {\it ci}
the partial derivative $A_X$ is non vanishing. Furthermore, since
\beq
A_X B_Y-B_X A_Y =2X
\label {ineq}
\enq
 has opposite signs for the two {\it ci}, we immediately find that the
 topological angles for the two  {\it ci} have opposite signs.
Therefore, circling around
the two {\it ci} gives a total accumulated phase of zero, rather than
$2\pi $,
 as might have been expected. This was indeed the  result found
computationally for large radius circling in some low lying $^2 A' $
states in
$C_2 H$  \cite {MebelYEB2001}. Thus we conclude that the above model is
suitable
for this molecule in the vicinity of the {\it ci}.

Example 2:

 In $C_{2v} $ symmetry the degeneracy between the $^2  B_2$ and $^2
A_1$
 states in $ Al - H_2$ \cite {Yarkony98}.
 Here symmetry arguments forbid the presence of an off diagonal term
of the form shown above in \er {AB1}. Noting further that
 the {\it ci} are in a plane orthogonal to the plane formed by the
three
atoms,
 we use $(X,Z)$ instead of $(X,Y)$ and write the off  diagonal term as
 $B(X,Z)$. Terms that are  made allowed have
the
form
\ber
A(X,Z) = (X^2 -1),~
B(X,Z) = XZ
\label {AB2}
\enr
The  reason that this off-diagonal term is allowed, is that $X$
transforms
 as $B_2$ and $Z$ as $A_1$.
 The two {\it ci} are located at $X ={\pm} 1 , Z=0 $. We now find for
the Jacobian
\beq
A_X B_Z-B_X A_Z =2X^2
\label {yineq}
\enq
and this gives the same sign for either {\it ci}.
The acquired phase is therefore the same and circling around both
yields 2$ \pi $.

The results of both Examples 1 and 2 have been previously obtained by
us, using
a different, graphical-algebraic procedure, the "continuous phase
tracing method"
\cite {EYBM2002}.

\subsection {Complex representation}
This representation is commonly used, e.g. for the doublet electronic
 states in the $E \otimes \epsilon$ Jahn-Teller
problem \cite {E72, ZwanzigerG}. The cartesian representation $(X,Y)$
of the
nuclear displacement coordinates is conveniently
replaced by the cylindrical polar coordinates $(q,\phi )$, according to
$X= q \cos \phi, Y=q \sin \phi $.
It has been shown \cite {EnglmanY2002} that when the vibronic coupling
(between nuclear and electronic
degrees of freedom) is taken to any arbitrary order, then the matrix in
\er {potential} (expressing this coupling) can be put in a complex
form,
 such that its off-diagonal part has the form
\beq
V_{12}(q,\phi )=Kqe^{-i\phi }[1+ q^{-2}\sum_{m=1...} q^{3m}
Q_{m+}(q)e^{3mi\phi} +
 \sum_{m=1...} q^{3m} Q_{m-}(q)e^{-3mi\phi}]
\label {offdiag}
\enq
Here K is a constant and $Q_{m+}$ and $Q_{m-}$ are polynomials in $q^2
$ with
 real coefficients that depend on the physical system (or of the model
used to
represent it) and whose leading term will be $q^0 $. (An alternative
expression in
a real representation is found in \cite {ThompsonM}.) Normally, for
stable
 physical systems, it is expected that with increasing $m$, $Q_{m+}$
and $Q_{m-}$
will both numerically decrease and so will, in each polynomial, the
coefficients
of successively higher powers of $q^2$.

There will also be diagonal, scalar and pseudo-scalar terms. The scalar
term
simply shifts the energies of the two states by an equal amount and
can be ignored. The pseudo-scalar term has the form
\beq
\sum_{m=1...} q^{3m} P_{m}(q)sin3m\phi
\enq
where the polynomials $P_m $ are defined in a similar manner to the
$Q$'s,
above. This term arises  when the system  is not time-reversal
invariant, such as in the presence of a magnetic
field. We shall not consider this situation. By consequence, the level
separation
is simply $2|V_{12}(q,\phi )|$ and the mixing angle is
\beq
\theta  =  {1\over 2} \arctan \frac{Im V_{12}}{Re V_{12}}
\label {theta2}
\enq
 Now, conically intersecting degeneracies
 of the two (complex) electronic states  can occur (those {\it ci} are
not the only ones possible)  at two types of points:
At $q=0$, coming from the initial, $q$-factor in \er {offdiag} and  at
trigonally
located degeneracies $(q_0, \phi_0 )$ from the square-bracket factor in
$V_{12}(q,\phi )$, such that one has either
\beq
\phi_0 =0, 2\pi /3, 4\pi /3
\label {trig1}
\enq
or
\beq
\phi_0 = \pi /3 , \pi , 5\pi /3
\label {trig2}
\enq

To find the sign of the phase acquired upon circling around any
trigonal
 {\it ci}, we use, as above,  the  circling radius $\Delta $ and the
 circling angle $\alpha $ shown in the figure. The polar coordinates,
 expressed in terms of these are, correct to the first order in $\Delta
$,
\beq
q  \approx  q_0 +\Delta \cos (\alpha -\phi_0),~~
\nonumber\\
\phi  \approx  \phi_0 +{\frac{\Delta }{q_0 }} \sin (\alpha -\phi_0)
\label {qphi}
\enq
We shall now find it convenient to rewrite \er {offdiag} in the form
\beq
V_{12}(q,\phi )=e^{-i\phi_0}U_{12}(q,\phi )
\label {offdiag1}
\enq
where $U_{12}(q,\phi )$ is defined by
\ber
U_{12}(q,\phi )=Kqe^{-i(\phi -\phi_0)}[1 &+& q^{-2}\sum_{m=1...} q^{3m}
Q_{m+}(q)e^{3mi\phi}
\nonumber \\
&+& \sum_{m=1...} q^{3m} Q_{m-}(q)e^{-3mi\phi}]
\label {offdiag3}
\enr
When this quantity is expanded about the {\it ci}, one obtains by \er
{qphi}
\beq
U_{12}( q, \phi ) \approx  U_{12} (q_0, \phi_0) + \Delta cos(\alpha -
\phi_0)
(\frac
{\partial U_{12}}{\partial q})_{q_0,\phi_0}+ \frac{\Delta}{q_0}
 sin(\alpha -\phi_0)(\frac
{\partial U_{12}}{\partial \phi})_{q_0,\phi_0}
\label {approxU}
\enq
The first term on the right vanishes at a point of degeneracy. Real and
imaginary parts of $U_{12} (q, \phi)$ come from the middle and the last
terms,
respectively. Therefore, by \er{theta2},
\beq
\theta  = -i\phi_0 + {1\over2} \arctan \Bigl(\frac{(\frac
{\partial U_{12}}{\partial \phi})_{q_0,\phi_0}}{iq_0 (\frac
{\partial U_{12}}{\partial q})_{q_0,\phi_0}} \tan (\alpha -
\phi_0)\Bigr)
\label {theta31}
\enq
The sign of the term linear in the  angle $\alpha -\phi_0$
 will determine the sign of the topological phase.

Example 3:
Let us choose, as in an earlier work \cite {EYACP}, a quartic
approximation
for the off-diagonal matrix element
\beq
V_{12}(q,\phi )=Kqe^{-i\phi }[1- \mu q e^{3i\phi } + \lambda q^3 e^{-
3i\phi }]
\label {offdiag2}
\enq
We first consider the case in which $\mu$ and $\lambda $ have the same
sign. Then, in
 addition to the zero at the origin, this expression can have up to
three {\it sets} of trigonally  positioned zeros,
 each of which gives rise to three {\it ci}. This is the maximum number
of zeros, but
 since a negative or complex $q_0 $ is physically not admissible, there
may
 be less than $3\times 3$ {\it ci}.

To determine the sign of the phases, one searches for the coefficient
in
 \er {theta31} of the linear term in $\alpha -\phi_0 $  as follows:
 At $q=0$, the coefficient is a constant: $-{1\over{2}} $.
 This gives, upon full circling, the usual topological phase of $-\pi$.
At a trigonal point, one finds for the prefactor of $\alpha -\phi_0$
\beq
{\frac {3\mu q_0 + 3\lambda q^3_0}{\mu q_0 -3\lambda q^3_0}}
\label {trigonal}
\enq
With the parameters $\lambda $ and $\mu$  having the same sign, the
above ratio
is positive or negative, depending on whether
\beq
\sqrt {\frac {\mu}{3\lambda }} > or < q_0 ~(the~radial~position~ of~
the~ {\it ci}~)
\label {criterion}
\enq
When equality holds, there is a double root and the intersection is no
longer
conical.

With the choices of $\mu =0.3, \lambda =0.003$ (used for illustration
in
 \cite {EYACP}), the following roots are found:

One root at $q=0$

Further, the following trigonal roots:
\ber
q_0=3.95, \phi_0 =0, 2\pi/3, 4\pi/3
\nonumber \\
q_0=7.42, \phi_0 =0, 2\pi/3, 4\pi/3
\nonumber \\
q_0=11.37, \phi_0= \pi, \pi/3, 5\pi/3
\label{roots}
\enr
 Substituting the parameter values in the left hand side of the
inequalities
in \er {criterion},
one finds for the square root the value of 5.77. This makes the
 signs of the phase of consecutive sets of {\it ci} come out
positive, negative and negative (in the order
of increasing radius), resulting in  accumulated phases of -$\pi $
(around the
origin only), then $2\pi $, then -$\pi $ and ultimately, at large
radius
circling, -$4\pi $. All of these values were confirmed independently by
the
 "continuous phase tracing method" \cite {EYACP} and by numerical
computation
 of the phase accumulated along the angular coordinate.
In the model of \cite {ZwanzigerG}, and of others (\cite {Yarkony98},
\cite {KoppelM} -\cite{KoizumiB})
identical to that one,  the off diagonal matrix element stops with the
quadratic
 term. In the language of our model, exhibited in \er {offdiag2}, this means that
the
coefficient $\lambda $ of the cubic term is zero. In virtue of the
criterion expressed
in \er{criterion} this implies a positive phase around each of the
three trigonal
{\it ci}, or for large-radius circling, when the central {\it ci} is
also circled,
a net phase change of +2$\pi $. This agrees with the results of the
cited
 works.

The  case of different signs in $\mu$ and $\lambda $ and also
$|\frac{\lambda^{\frac{1}{2}}}
{2\mu^{\frac{3}{2}}}| <1 $ yields three new type of roots,
 as follows:
\beq
q_0 = \sqrt{\frac{-\mu}{\lambda}}
\label{qroot}
\enq
required to be real and positive, with angles $\phi_0$ at the
 intersection that are
\beq
{1\over3}\arccos[\sqrt{\frac{\lambda}{-\mu}}/(2\mu)]
\label{phiroot}
\enq
in which the argument of $\arccos$ is required to be magnitude-wise
less than one, as well as
 two further angles oriented with respect to the previous by
$2\pi /3 $ and $4\pi /3 $.  The roots can be checked upon substitution
into
 \er {offdiag2}.

The signs of the phase angles can be obtained from
 \er{theta2} and \er {approxU}. After some manipulations one obtains an
expression,
 which reads, correct to the first power of $tan(\alpha - \phi_0)$,
\beq
\theta =-{1\over 2}\arctan[ {1\over 2}\tan(3\phi_0) +\frac{3}{4}
(\tan(3\phi_0))^2
tan(\alpha - \phi_0)]
\label{theta3}
\enq
Since  $\arctan$  is an increasing function of its argument, and noting
the negative
sign in the expression, we find that circling  around these {\it ci}
gives
 a negative phase angle.
This last type of {\it ci}, which is also trigonally located  but
angularly
 shifted from the positions given
 in \er {trig1} or \er{trig2}, has not been noted before, neither
experimentally,
nor theoretically or calculationally.

\subsection{Phases in excited states}
\noindent
(a) A three dimensional parameter space.

The following Hamiltonian was used by Berry \cite {Berry84} for
elucidation of
the geometric phase in a three-dimensional parameter space
represented by ${\bf R} =(X,Y,Z)$. The same
 Hamiltonian formed the basis of discussion of a magnetic monopole in
real space  \cite {WuYang}.
\beq
H = {1\over 2}\left( \begin{array}{cc}
  Z  &  X-iY \\
  X+iY  &  -Z
\end{array} \right)
\label{H1}
\enq
This can also be written in terms of Pauli spin matrices, in the form
\beq
H ={1\over2} (X\sigma_x + Y\sigma_y + Z\sigma_z)
\label{H2}
\enq
The (adiabatic) eigenstates of the Hamiltonian can be obtained in terms
of the
 spherical coordinate representation of ${\bf R} = (R,\theta,\phi) $ as
\beq
\psi_l =\left( \begin{array}{cc}
  -sin\frac{\theta}{2} e^{-i\phi /2}  \\
  cos\frac{\theta}{2} e^{i\phi /2}
\end{array} \right)
\label{psil}
\enq
for the lower state, and
\beq
\psi_u = \left( \begin{array}{cc}
  cos\frac{\theta}{2} e^{-i\phi /2}  \\
  sin\frac{\theta}{2} e^{i\phi /2}
\end{array} \right)
\label{psiu}
\enq
for the upper (excited) state. The separation of the states is $\Delta
= R$.
It was shown, in e.g. \cite{Child}, that the geometric angle $\gamma_C$
arising
from describing a closed contour $C$ in the $\bf R$- (or parameter) -
space
can be expressed in terms of the  {\it expectation value} of gradient
of the
  Hamiltonian
\beq
\Delta_{\bf R} H= {1\over 2}({\bf {\hat X}} \sigma_x + {\bf {\hat Y}}
\sigma_y +
 {\bf{\hat Z}} \sigma_z)
\label{delta}
\enq
namely
\beq
\gamma_C (\psi) = \int_S \frac{(\psi|\Delta_{\bf R} H|\psi)}{R^2}\cdot
d{\bf S}
\label{gamma2}
\enq
where the integral is taken over a signed area $\bf S$ enclosed by the
contour
$C$. Evaluating the expectation values and carrying out the surface
integration
for a contour at constant $\theta$ (namely, over a region enclosing a
spherical cap), one obtains the following geometric phases for the two
eigenstates.
\beq
\gamma_C (\psi_l) = -(1-\cos\theta )\pi
\label {gamma3}
\enq
for the lower (ground) state, and its opposite
\beq
\gamma_C (\psi_u) = (1-\cos\theta )\pi
\label {gamma4}
\enq
for the upper (excited) state. One sees that, for a contour around the
large
circle, for which  $\theta = {1\over2} \pi$, the
 phase-factors,
namely $e^{i\gamma_C}$,  are the same $(-1)$ for the two states, though
{\it not}
the phase-angles, whereas for other values of the angle $\theta$ not
even the
phase factors are the same. In principle, phase factor differences can
be established
experimentally by interference measurements.

(b) Adiabatic (slow) time development in a doublet.

 We consider the time-dependent {\SE} written as
\beq
i\frac {\partial \Psi(x,t)}{\partial t}  = H(x,t)\Psi(x,t)
\label {TDSE}
\enq
(in which $t$ is time, $x$ denotes all particle coordinates, $H(x,t)$
is a
 real time dependent Hamiltonian and $\hbar =1$). As is well known,
 the presence of $i$ in  the equation causes
 the solution $\Psi (x,t)$  to be complex-valued. We now consider a
special case
for the Hamiltonians in which two externally imposed independent
sinusoidal
 perturbation
interact linearly with the electron. Further, the electronic Hilbert
space is
confined to a degenerate electronic doublet, represented by two
vectors:{\tiny
 $ \left(\begin{array}{c}
 1 \\ 0  \end{array} \right) $} and {\tiny $ \left(\begin{array}{c}
 0 \\ 1  \end{array} \right) $}. In this representation
the Hamiltonian can be written as a 2x2 matrix, for
which we have chosen the following form, whose physical origin
has been  described in various works (e.g., \cite {MooreStedman} -
\cite {EYB1})
\beq
H(t)= G/2\left( \begin{array}{cc}
  -\cos(\omega t) &  \sin(\omega t) \\
  \sin(\omega t)  &  \cos(\omega t)
\end{array} \right)
\label{h}
\enq
Here $\omega $ is the angular frequency of the two external
disturbances, taken to be the same
for both. The eigenvalues of \er{h} are $-\frac{G}{2}$ and
$\frac{G}{2}$. We
 take  $G>0$, so that the former is that ground state energy. A method
of solution was outlined in \cite {EYB1}
and the amplitude of the vector  {\tiny $ \left(\begin{array}{c}
 1 \\ 0  \end{array} \right) $}   in the ground state was given there
by
 the expression
\ber
\chi_1 ^g (t) =  \cos (Kt) \cos ( \omega t/2) + (\omega/2K) \sin( K t)
\sin (\omega t/2)
\nonumber\\
  +i(G/2K) \sin(K t) \cos( \omega t/2)
\label{CGspec}
\enr
with
\beq
K =0.5 \sqrt{G^2 + \omega^2} \approx G/2
\label {Kexpr}
\enq
the latter approximation being justified in the adiabatic, slow motion
limit
defined by
\beq
\qquad G/\omega \gg 1
\label {adiab}
\enq
In the present work we do not repeat the method of solution  in \cite
{EYB1},
 but use it to present the results also for the  amplitude of
 the vector  {\tiny $ \left(\begin{array}{c}
 0 \\ 1  \end{array} \right) $} for the ground state to be denoted by
${\chi_2 ^g (t)}$, as well as the corresponding amplitudes ${\chi_1 ^e
(t)}$
and ${\chi_2 ^e (t) }$ in the excited state. The two states are
differentiated
by the initial conditions, for $t=0$:
\beq
\chi_1 ^g (0) = 1, \chi_2 ^g (0) = 0
\label {grt0}
\enq
for the ground state and
\beq
\chi_1 ^e (0) = 0, \chi_2 ^e (0) = 1
\label {ext0}
\enq
 for the excited state. We recall the well known result that in the
adiabatic
limit the state  maintains its ground or excited state character
throughout
the motion \cite {BornF}. After some algebra one obtains the following
result,
which encapsulates the four amplitudes in one single formula for
arbitrary
initial values $\chi_1 (0)$ and $\chi_2 (0)$:
\ber
 \chi_{{\tiny
  \left(\begin{array}{c}
 + \\ -  \end{array} \right) }} (t)  &=& {\tiny
 \left(\begin{array}{c}
 1 \\-i \end{array} \right)}\frac{1}{4K}\lbrace  e^{i(K-{1\over2}\omega
)t}
[\chi_1 (0)(K+ \frac{G}{2}+\frac{\omega}{2})(1 {\tiny
  \left(\begin{array}{c}
 + \\ -  \end{array} \right) }f_1 e^{i\omega t})
\nonumber\\
&+& i\chi_2 (0)(K-\frac{G}{2}+\frac{\omega}{2})(1 {\tiny
  \left(\begin{array}{c}
 - \\ +  \end{array} \right) }f_2 e^{
i\omega t})]
\nonumber\\
& {\tiny \left(\begin{array}{c} + \\ -  \end{array} \right)}&
e^{-i(K-{1\over2}\omega )t}[\chi_1 (0)(K-
\frac{G}{2}+\frac{\omega}{2})
(1 {\tiny  \left(\begin{array}{c} + \\ -  \end{array} \right) }f_2 e^{-i\omega t})
\nonumber\\
&-& i\chi_2 (0)(K+\frac{G}{2}+\frac{\omega}{2})(1 {\tiny
 \left(\begin{array}{c}
 - \\ +  \end{array} \right) }f_1 e^{-i\omega t})] \rbrace
\label {chiexp}
\enr
where the upper and lower symbols in the parentheses have to be taken consistently with
the
 choice of the components. The formula, as presented, is valid
generally, for
arbitrary boundary conditions. For the special choice of either the
ground
or the excited state the initial conditions, as expressed in
\er {grt0} or in \er {ext0}, have to be substituted in the above
formula.
The auxiliary functions $f_1$ and $f_2$ are given by
\beq
f_1= {{K+\frac{G}{2} -\frac{\omega}{2}} \over{K+\frac{G}{2}
+\frac{\omega}{2} }}, ~~~
f_2= {{K-\frac{G}{2} -\frac{\omega}{2}} \over{K-\frac{G}{2}
+\frac{\omega}{2} }}
\label{fs}
\enq

In the adiabatic limit, when $G/\omega \gg 1$ and therefore
 $K ={1\over2} \sqrt{G^2 + \omega^2} \approx G/2$  correct to the first
order
in $\frac{\omega}{G}$, the  expression simplifies to take the form:
\ber
 \chi_{{\tiny \left(\begin{array}{c} + \\- \end{array} \right) }} (t)  &=&
 {\tiny \left(\begin{array}{c}1 \\ -i  \end{array} \right) }
 \frac{1}{2}\lbrace  e^{i(G-{1\over2}\omega)t}
[\chi_1 (0)(1+\frac{\omega}{2G})(1 {\tiny
  \left(\begin{array}{c} + \\ -  \end{array} \right) }f_1 e^{i\omega t})
\nonumber\\
&+& i\chi_2 (0)(\frac{\omega}{2G})(1 {\tiny
  \left(\begin{array}{c} - \\ +  \end{array} \right) }f_2 e^{i\omega t})]
\nonumber\\
&{{\tiny \left(\begin{array}{c} + \\- \end{array} \right) }}&
e^{-i(G-{1\over2}\omega)t}[\chi_1 (0)(\frac{\omega}{2G})
(1 {\tiny \left(\begin{array}{c} + \\ -  \end{array} \right) }f_2 e^{-i\omega t})
\nonumber\\
&-& i\chi_2 (0)(1+\frac{\omega}{2G})(1 {\tiny
  \left(\begin{array}{c} - \\ +  \end{array} \right) }f_1 e^{-i\omega t})]\rbrace
\label {chiexpad}
\enr

The crucial element in the establishment of the proper sign of
 the topological phase is the recognition that the prefactors $f_1$ and
$f_2$ of the circular functions $e^{\pm i\omega t}$ in \er{fs} are
magnitude-wise
 less than unity. Therefore, upon circling
a full period, none of the round brackets containing these in the above
 expression can cause a sign change.  Thus the sign of the phase is
dictated
 by whether the first or
the second term in \er {chiexpad} stays finite (non-vanishing) in the
 adiabatic limit $\frac{\omega }{G} \to 0 $.
It is now evident from the initial conditions in \er {grt0} that
for the ground state it is the first
term that stays finite, and therefore the topological phase is $-\pi$,
 whereas for the excited state, for which the initial conditions are
\er {ext0},
 it is the second term that survives and yields
the topological phase of $+\pi$. Note also, that the first exponential
factors
$ e^{{\pm}iGt} $ arise from the dynamical phase and are irrelevant to
 the topological phase in the adiabatic limit. For their effect upon
the topological
 phase in the not fully adiabatic case, we refer to \cite {EYB2}. We
further
 find from the expression that, in either the ground or the excited
state,
 both components  $1$ and $2$
(diferentiated in the above expressions by round and square
brackets, respectively) have the same topological phase.

Thus, in summary, we find that the signs of the topological phases are
identical
for both components in a given state, but are opposite in the ground
and excited states. It is possible to interpret this result from a
 wider perspective, as follows:

The time-dependent ground state of a bound system is characterized by a
 {\it lower} boundedness of the energy. This leads to the analyticity
of the
wave function (more precisely of its logarithm) in the lower half of
 the complex t-plane \cite {Khalfin, PerelmanE}. The phase is then
obtained
 from the modulus by a Kramers-Kronig type relation, which  involves
integration along a contour consisting of the real t-axis and a
large semicircle in the lower half of the t-plane \cite{EY1,
EYB1, EYB2}. In contrast, the upper partner in a doublet has an energy
{\it upper}
boundedness. This leads to an analyticity in the upper half t-plane, to
a closing
of contours (in the Kramers-Kronig relation) in the upper half plane
and to
 a  topological phase of the opposite sign.

In a general situation that involves in the Born-Oppenheimer
superposition
more than two electronic states, one still expects a negatively signed
topological phase in the adiabatic ground state, but in general one
cannot
 predict the sign in an intermediate-energy adiabatic state.

\section {Pseudo-fields Arising from Degeneracies}

Denoting by $|i>$ a set of electronic adiabatic eigenfunctions  and by ${\bf \vec \nabla}$
the gradient operator conjugate to the nuclear coordinates, we introduce the non-adiabatic
coupling terms (NACTs) as
\beq
\tau_{ij} =<i|{\bf \vec \nabla}|j>
\label{taudef}
\enq
It was noted  some time ago that  the NACTs
 can be incorporated in the nuclear part of the \SE as
a
vector potential ${\bf A }$ \cite {Smith}- \cite {MeadT}. The
question of
a possible magnetic field $ {\bf H}$ has been considered in \cite
{Baer2001}
. This field is not due to any source external to the
molecule, but rather arises from the mutual coupling between the
electrons
and the nuclei. To mark this point we call $\bf H$ a pseudo-field.
Formally,
it is associated with ${\bf A }$  through
\beq
{\bf H} = curl\bf A
\label {H}
\enq
 In the present work we
derive the magnetic field $\bf H$, for the case where the electron-
nucleus
coupling is expected to have
an especial importance, namely, for two (diabatic) electronic states
becoming degenerate at a single point in a two dimensional ($2D$)
parameter
space. In the context of a polyatomic molecule (which is our subject of
reference), this $2D$ space is the plane created by a pair of nuclear
displacement modes, to be designated $X$ and $Y$. The degeneracy of the
electronic states gives rise to a  {\it ci},
 conventionally regarded as the origin $X=Y=0$. The NACTs have a
 pole at the origin and, excluding this point of singularity but still
 staying in the neighborhood of the {\it ci} , the curl of the NACTs
vanishes \cite{Baer75,BaerPR}.
 Farther away from the {\it ci} , the curl of the NACTs is non-
vanishing, provided some {\it ci} due to other adiabatic states are found
in the region. These two
facts (the "vanishing of the curl" around the {\it ci} and the existence
of a pole at the origin) cause
the
issue of the magnetic field to be problematic. The approach in this
work is
to obtain the field associated with the {\it ci} by a limiting
procedure,
which yields the field unambiguously.

 The magnetic field that we derive is tied to the fact that in the
neighborhood  of the {\it ci} we have an electronic sub-manifold of at
least
two dimensions. (The actual electronic Hilbert space is, of course, of
higher dimensionality,
 but this higher dimensionality can be ignored near the {\it ci} ,
whereas a one
dimesional space cannot give rise to a {\it ci}, by
definition.)
 However, as will be verified in the sequel, the NACT-induced magnetic
field
 will vary with the kind of superposition one makes of the electronic
states
 within the sub-manifold. In the following section we first obtain the
 magnetic field in the representation of the adiabatic states.  These
are
 the appropriate choices under slow variation of nuclear coordinates,
since then the adiabatic states are eigen-states which are
uncoupled from the rest of the manifold. Subsequently, we shall also
 consider a magnetic field in another (non-adiabatic) representation.
The representation dependence of the magnetic field is connected to
the non-Abelian nature of the
situation.
 ("Non-Abelian" means the involvement of more than one electronic state
and
 the presence of non commuting matrices in the Hamiltonian. The need
to
specify the representation does
 not arise in Abelian (commuting) systems, which is the usual
background
 for classical electromagnetism. There one defines a unique magnetic
field
without regard to the electronic state the system is in. The magnetic
 field is, trivially, "unique" for a single electronic state.
 However, it is such  also for a many state situation, such that
 the Hamiltonian contains no non-commuting matrices.)

 Another objective
of this work is to obtain, in addition to the magnetic field, the
tensorial Yang-
Mills
 field. By its original conception \cite {YangM} and its numerous later
developments and applications (e.g., \cite{Berry84} and \cite {Weinberg}), the Yang-Mills
field
 constitutes a residual interaction within the manifold  (in our
context,
between the electronic states).
From a historical perspective, there is a long and distinguished line of papers
 in which a magnetic
field has been obtained for a related problem: a pair of states that
become
 degenerate at a point in a three dimensional parameter space.
Notable among these works  are Dirac's derivations of the quantized monopole-
field
\cite{Dirac31}, Wu and Yang's matched vector potential \cite {WuY} and
Berry's formulae for the magnetic field (e.g., \cite {Berry84}, \cite
{Berry90},
  as are also other papers reprinted  in Shapere and Wilczek's volume
\cite
 {ShapereW}.)

\subsection {The adiabatic states}

We start with the following Hamiltonian (taken over from Berry's ground
 laying paper \cite {Berry84} with one essential modification)
\beq
H_b  = \left(\begin{array}{cc}
           bZ & X-iY\\
           X+iY & -bZ
\end{array} \right)\qquad\\
\label {BerryH}
\enq

This is written in a diabatic representation of the electronic states
(that
 are such that the state does not involve, at least locally, nuclear
 coordinates). A Hamiltonian matrix written out in a different representation,
 employed, e.g., in \cite {Baer2001} but formally different from that in
 \er{BerryH} is considered later in this paper.
 The two components involved in the formalism are denoted
by
$ |+>, |-> $, and the nuclear mode coordinates by $(X,Y,Z) (={\bf R})$.
The parameter b in \er {BerryH} is to be noted. The magnetic monopole
treatment (e.g., in \cite {Berry84, WuY}) has $b=1$ and "spherical
symmetry".
 We shall obtain the NACT and a magnetic field for a general $b\ne 1$,
and
then proceed to the $ b\to 0$ limit, which is the situation of interest
 to us, namely the two parameter problem. (The parameter $Z$ varies
along the
 so-called seam-direction of the degeneracies. In the limit of $b=0$,
$Z$
 does not enter the dynamics. Physically, $bZ$ represents a magnetic field
 aligned with the quantization direction of the two states in \er{BerryH}.
 We write it as $\pm bZ$ and then let $b \to 0$, rather than let
 $Z\to 0$, first to maintain the analogy with \cite {Berry84}, and also
  since a coordinate $Z$ cannot be made to vanish in a Hamiltonian.)
   The matrix in \er {BerryH} has
eigenvalues
  $\pm  R_b$ and the corresponding adiabatic states take the form
(for $b>0$):
\beq
|1> =  e^{-i\phi /2} \cos(\theta_b /2)|+> + e^{i\phi /2} \sin(\theta_b
/2)|->\\
\label {1adstate}
\enq
\beq
|2> = -e^{-i\phi /2} \sin (\theta_b /2) |+> + e^{i\phi /2}
\cos(\theta_b /2)|->
\label {2adstate}
\enq
where
\beq
R_b =\sqrt{X^2 +Y^2+(bZ)^2} =\sqrt {q^2+(bZ)^2}
\label {Rb}
\enq
\beq
\tan\phi  = Y/X \\
\label {phi}
\enq
\beq
\tan \theta_b =q/bZ
\label {theta}
\enq
having introduced  the radial coordinate $q$ in the cylindrical polar
 description $(q,\phi )$ through
\beq
q^2=X^2+Y^2
\label {q}
\enq

The NACTs  have already been defined as elements of the matrix (denoted by ${\bf
\tau} ^b$)
  of the gradient operator ${\bf \vec \nabla} $   conjugate to the
variables $(X,Y,Z)$.
Thus ${\bf \tau} ^b$ is both a vector and a matrix. Its elements in the
adiabatic-state representation, \er {1adstate}- \er {2adstate}, are:
\ber
{\bf\tau} ^b_{11} & \equiv & <1|{\bf \vec \nabla}|1> = -{1\over 2}i
\cos \theta _b
{\bf \vec \nabla} \phi
\label {tau11f}\\
                & =&-{\bf \tau}^b_{22}
\label{tau11}
\enr
\beq
{\bf \tau} ^b_{12}  = {1\over 2}i \sin \theta _b {\bf \vec \nabla} \phi
-{1\over2} {\bf \vec \nabla} \theta _b
\label {tau12}
\enq
\beq
{\bf\tau}^b_{21}  = {1\over 2}i \sin \theta _b {\bf \vec \nabla} \phi+
{1\over2}{\bf \vec \nabla} \theta _b
\label {tau21}
\enq

[This is a departure from the usual definition in the chemistry
literature of
 NACTs, which are real off-diagonal and anti-symmetric matrices (or
matrix
 elements). However, we shall find that our diagonal  NACT are
anti-hermitean, just as are the usual NACTs. (Trivially, i times the
NACTs
 are hermitean. The usual definitions of the NACTs can be regained by
redefining, say, the second adiabatic state states to be i times the
ones
used here. Our choice makes the "off-diagonal" magnetic field, which
can be
seen below in (16) and (18), come out real.)
Particle physicists use the term connectivity ($\bf A$), which also is
a
matrix-vector, related to $\bf\tau$  by
\beq
 {\bf A} =i{\bf\tau}
\label{A}
\enq
 (as in \cite {Jackiw}). They further employ the terms curvature (or
the
 Yang-Mills field intensity tensor), defined by
\beq
{\bf F} = {\bf \vec \nabla} \wedge {\bf A}  - i {\bf A} \wedge {\bf A}
=
{\bf H} - i {\bf A} \wedge {\bf A}
\label {YMF}
\enq
and the Berry-phase or line integral over a closed contour C
\beq
\int_C {\bf A} \cdot d {\bf R}
\label{integral}
\enq
(The conditions, in a molecular context, for the vanishing of {\bf F} have been
described in \cite {Baer75}.)

The magnetic field $\bf H$ arising from the NACTs is
\beq
{\bf H}=i {\bf \vec \nabla} \wedge {\bf \tau}
\label {Htaudef}
\enq
We first calculate this in the representation of the adiabatic states
shown in \er{1adstate}-\er{2adstate}.
\ber
{\bf H}^b_{11} &=& -{\bf H}^b_{22}=i {\bf \nabla} \wedge {\bf
\tau}^b_{11}\\
&=& {1 \over 2} {\bf \vec \nabla} \wedge(\cos \theta_b {\bf \vec
\nabla} \phi) \\
&=& -{1 \over 2}\sin \theta_b ({\bf \vec \nabla} \theta_b \wedge {\bf
\vec \nabla}\phi)
   +{1\over2}\cos\theta_b ({\bf \vec \nabla} \wedge {\bf \vec
\nabla}\phi)
\label {H11}
\enr
and
\ber
{\bf H}^b_{12} &=& {\bf H}^b_{21}=i{\bf \vec \nabla} \wedge {\bf
\tau}^b_{12}\\
&=& -{1 \over 2} {\bf \vec \nabla} \wedge(\sin \theta_b {\bf \vec
\nabla} \phi)\\
&=& -{1 \over 2} \cos \theta_b ({\bf \vec \nabla} \theta_b \wedge {\bf
\vec \nabla}\phi)
   -{1\over2}\sin\theta_b ({\bf \vec \nabla} \wedge {\bf \vec
\nabla}\phi)
\label {H12}
\enr
having put ${\bf \vec \nabla}\wedge {\bf \vec \nabla}\theta_b=0$ by
elementary vector algebra.
The (apparently) similar quantity ${\bf \vec \nabla} \wedge {\bf \vec
\nabla}\phi$ needs a
different treatment, because of the singular nature of $\phi$ on the
seam
line $q=0$. This will be done below.

We next calculate the Yang-Mills (tensorial) field, given by
\beq
{\bf F}=i({\bf \vec \nabla} \wedge {\bf \tau} +{\bf \tau} \wedge {\bf
\tau})
= {\bf H} +i {\bf \tau} \wedge {\bf \tau}
\label {YMF1}
\enq
also in the representation of the adiabatic states
\ber
{\bf F}^b_{11} &=& {\bf H}^b_{11}+i{\bf \tau}^b_{12} \wedge {\bf
\tau}^b_{21}\\
&=& {1\over 2} {\bf \vec \nabla}\wedge (\cos \theta_b {\bf \vec
\nabla}\phi)+
{1 \over 2} \sin \theta_b ({\bf \vec \nabla} \theta_b \wedge {\bf \vec
\nabla}\phi)\\
&=& -{1 \over 2} \sin \theta_b ({\bf \vec \nabla}\theta_b \wedge {\bf
\vec \nabla}\phi)
 +{1\over 2} \cos\theta_b ({\bf \vec \nabla}\wedge {\bf \vec
\nabla}\phi)
\nonumber\\
&+& {1\over 2}\sin \theta_b ({\bf \vec \nabla} \theta_b \wedge {\bf \vec
\nabla}\phi)
\label{F11inter}\\
&=& {1\over2}\cos\theta_b ({\bf \vec \nabla}\wedge {\bf \vec
\nabla}\phi)
\label{F11next}\\
&=& - {\bf F}^b_{22}
\label {F11}
\enr
and similarly
\ber
{\bf F}^b_{12} &=& {\bf H}^b_{12}+i({\bf \tau}^b_{11} \wedge {\bf
\tau}^b_{12} + {\bf \tau}^b_{12} \wedge {\bf \tau}^b_{22})
\\
&=& -{1 \over 2} {\bf \vec \nabla} \wedge(\sin \theta_b {\bf \vec
\nabla} \phi)+ {1\over 2} \cos \theta_b ({\bf \vec \nabla} \theta_b \wedge {\bf \vec
\nabla}\phi)
\\
&=& {1\over 2}\cos \theta_b ({\bf \vec \nabla} \theta_b \wedge {\bf
\vec \nabla}\phi) -{1\over 2}\sin\theta_b ({\bf \vec  \nabla} \wedge {\bf \vec
\nabla}\phi)
\nonumber \\
 &-& {1\over 2}\cos \theta_b ({\bf \vec  \nabla} \theta_b \wedge {\bf \vec
\nabla}\phi)
\label{F12inter}
\\
&=&- {1\over2}\sin\theta_b ({\bf \vec \nabla}\wedge {\bf \vec
\nabla}\phi)
\label{F12next}
\\
&=&  {\bf F}^b_{21}
\label {F12}
\enr
In the above expression we call attention to the cancellation between
 the terms
in the curl and in the vector product (the first and the third terms
in
\er{F11inter} and \er{F12inter})

The derivative quantities ${\bf\vec \nabla}\phi$ in the NACTs
(\er{tau11} and \er{tau12})
 and  ${\bf \vec \nabla} \wedge {\bf \vec \nabla}\phi$ in the field
intensities
(\er{F11next} and \er{F12next}) do not depend on $b$. They have been
 considered in
\cite {MeadT}, section III.C in a somewhat different context, namely as
arising
from a "pure-gauge" phase factor that multiplies the {\it whole}
adiabatic
state. Their conclusion is that they give rise to a pseudo-magnetic
field
that is zero everywhere except along the "curve of intersection [the
seam] of
the two potential surfaces ...where it has a delta-function
singularity".
 We shall use this result in the formulae that immediately  follow,
but note that quantities that depend on $b$ are new to the present work
and
 so are some results that do not vanish in the limit of $b \to 0$. (A
formal
justification of the result of \cite {MeadT} involves the extension of
Stokes' theorem to singular integrands and will be given elsewhere
\cite {YahalomE}.)
 The $\phi$-derivatives
will now be given  in two coordinate systems: 1) a cylindrical
coordinate
system $(q,\phi,Z)$ with {\bf unit
vectors} depicted  by bold and hatted symbols and 2) in the Cartesian
space
 $(X,Y,Z)$, for which we
quote the results of \cite {MeadT}, using the notation of $({\bf i,j,k}
)$ for the unit
orthogonal vectors.
\beq
{\bf \vec \nabla} \phi = {\bf \hat \phi} /q ~ [= (X{\bf j} - Y{\bf i}
)/(X^2 + Y^2)]
\label{nablaphi}
\enq
\beq
{\bf \vec \nabla} \wedge {\bf \vec \nabla}\phi = {\bf\hat Z}\delta(q)/q
~ [=2\pi \delta(X) \delta
(Y){\bf k}]
\label {curldivphi}
\enq
$\delta(q)$ is the Dirac delta function. This result is in accord with
\cite {Baer75}.

It thus immediately follows that the tensorial fields shown in  \er{F11}
and \er{F12} are zero, except on the seam, irrespective of the value of $b$/

\subsection{The Non Adiabatic Coupling Terms (NACTs)}

The $\sin\theta_b$ and $\cos\theta_b$ prefactors of these derivatives
depend on
 the parameter $b$, as the notation indicates, and  the results are
 true for any, general value of $b$. However, we shall
 examine what happens in the $b\to 0$ limit, namely as the 2-parameter
(or
2-dimensional) problem is reached. We recall that the angle $\theta_b$
has been defined in \er{theta}. From this definition, in the limit
$b\to 0$,
 we  have, for $ q \ne 0$,
\beq
\lim_{b\to 0}\sin\theta_b (= q/R_b) \to 1~~  (\sin(\frac{\theta_b}{2})
=\cos(\frac{\theta_b}{2})=\frac{1}{\sqrt2})
\label{sinlimit}
\enq
Clearly, also
\beq
\lim_{b\to 0}  \cos\theta_b (= bZ/R_b) \to \pi \delta (q) R_b \to \pi
\delta (q) q
\label{coslimit}
\enq

To obtain the last result we have used the expression for the delta
function
\beq
\pi \delta (q)\ =\lim_{\epsilon \to 0} \frac{\epsilon}{(\epsilon^2 +
q^2)}
\label{deltaofdirac}
\enq
and the definition of $R_b$ in \er{Rb}. We add two further results to
be used in
 the sequel. The first is
\beq
{\bf \vec \nabla} \theta_b = -(bq/R_b^2) {\bf\hat Z}
+(bZ/R_b^2){\bf\hat q}~
 [= -(bq/R_b^2){\bf k}  +(bZ/qR_b^2)(X{\bf i}+Y {\bf j})]
\label {nablatheta}
\enq
and from this,  using \er {nablaphi},
\ber
{\bf\vec \nabla} \theta_b \wedge {\bf\vec \nabla} \phi &=&
(b/R_b^2){\bf\hat q}
 +(bZ/q R_b^2){\bf\hat Z}\\
&[=&(b/qR_b^2)(X{\bf i} + Y{\bf j}) +(bZ/qR_b^2){\bf k} ]\\
&=& (bZ / R_b^2) ({\bf \hat q} /Z +{\bf \hat Z} /q)
\label{nablanabla}
\enr

Then, from \er {tau11f} and \er{nablaphi},
\ber
{\bf\tau}^b_{11} & =& -{i\over 2} (bZ/qR_b){\bf\hat\phi}\\
&[=& -{i\over 2}(bZ/R_b)(X{\bf j} - Y{\bf i} )/(X^2 + Y^2)]\\
& =&-{\bf\tau} ^b_{22}
\label{tau11ad}
\enr
\ber
{\bf \tau}^b_{12} & =& {i\over 2}\frac{1}{R_b}{\bf\hat\phi} -
\frac{bZq}{R_b^2}(\frac{{\bf\hat q}}{q}
 -\frac{{\bf\hat Z}}{Z})\\
{\bf\tau} ^b_{21} & =& {i\over 2}\frac{1}{R_b}{\bf\hat\phi}
+\frac{bZq}{R_b^2}(\frac{{\bf\hat q}}{q}
 -\frac{{\bf\hat Z}}{Z})
\label {tau12ad}
\enr

From  \er{coslimit} we see immediately that (because of the delta
function
 factor arising from $bZ/R_b$) the diagonal NACTs are zero outside the
 seam line $q=0$ as $b\to 0$ .
 However, on the seam line they are non-zero. This is a new result,
which
has been obtained
 by going to the limit $b \to 0$ after calculation of the derivatives.

If we now calculate the Berry-phase or the line integral shown in \er
{integral}
 taken with a  circular contour encircling the seam line at any finite
 distance $(q>0)$ from it, we obtain
\beq
\int_C {\bf A}_{11} \cdot d {\bf R}= \pi \cos \theta_b
\label{integral11}
\enq

In the 2-parameter limit,  as $b\to 0$, this will tend to zero (from
\er{coslimit}).
 By familiar
 arguments (based on Stokes' theorem that equate  the contour integral
with
a surface integral of the curl), the flux across any finite part of the
 parameters plane of the "diagonal" magnetic field in the adiabatic
 representation vanishes, notwithstanding the fact that the vector
potential
 is not zero inside the (infinitely thin) $q=0$ "solenoid". (We shall
presently
 check this result by actually calculating the surface integral of the
 "diagonal" magnetic field.)
    In the off-diagonal NACT given in \er{tau12} or \er{tau21}, the
 second term can
be seen [from its form in \er{nablatheta}] to give $\delta (q)q$ as $b
\to 0$,
whose line integral around a finite circle $q>0$ clearly vanishes.
However, the first term in the off-diagonal NACTs survives the $b \to
0$ limit
even for $q>0$. Thus we have a genuine, finite vector
potential entering through the off-diagonal part. The resulting line
integral
 \beq
\int_C {\bf A_{12}} \cdot d {\bf R}=- \pi \sin \theta_b
\label{integral12}
\enq
is responsible for the Berry phase of -$\pi$.

\subsection {The magnetic field in the adiabatic representation}

As already noted above, we can check the previous line-inegral results
by
 the equivalent method of evaluating the flux of
 the magnetic field $\bf H$ across an $XY$- plane. $\bf H$ can be
calculated
 from \er{H11} and from \er{H12}. For the diagonal part we obtain
\ber
{\bf H}^b_{11}&=& [(bZ/2qR_b)\delta (q) - (bZ/2R_b^2)]{\bf\hat Z} -
 (bq/2R_b^2){\bf\hat q}\\
&[=&(\pi (bZ/R_b)\delta (X)\delta (Y)- (bZ/2R_b^3)){\bf k}
\nonumber\\
&&-(b/2R_b^2)(X{\bf i}
+Y{\bf j})]\\
&=&- {\bf H}^b_{22}
\label {H11ad}
\enr
The second expression (in square brackets) is written in the Cartesian
frame.
In the $b\to 0$ limit this field vanishes outside the seam line $q=0$.
In the
general case, for which $b>0$
 the last term is oriented along the cylindrical radius vector $({\bf
q})$
 and only the $Z$
 component of the field contributes to the flux across the $(q,\phi)$
[or
 the $(X,Y)$]-plane. Corresponding to the result for a circular contour
 at  $q$ shown in \er{integral11} which vanishes in the $b \to 0$ limit
(as seen
 above), one obtains after some manipulation of the surface integral
\ber
 \lim _{b\to 0 }(Magnetic \ flux)_{11} &=& \pi\int_0 ^{\infty} dq \delta (q)  - \pi
bZ\int_{0}^{\infty}
 dq \frac{q}{ R_b^3} \\
 &=& 0\\
&=& -(Magnetic flux)_{22}
\label{integral11ad}
\enr
Thus the two methods  (the line integral and the surface integration)
give the same result.

    The off-diagonal magnetic field is from \er{H12}
\ber
{\bf H}^b_{12}&=&-{1\over2} (\frac{\delta
(q)}{R_b}+\frac{(bZ)^2}{qR_b^3})
{\bf\hat Z} -
 {1\over2}(b^2 Z/R_b^3){\bf\hat q}\\
&[=&(-\pi (q/R_b)\delta (X)\delta (Y)- (bZ)^2/2qR_b^3){\bf k}\\
&& -(b^2Z/2qR_b^3)(X{\bf i}+Y{\bf j})]\\
&=& {\bf H}^b_{21}
\label {H121ad}
\enr

In the limit $b \to 0$, this field also vanishes outside q=0. To
evaluate the
 flux,  care must be taken to go to the limit $b\to 0$ only after the
integrations are
 performed. At the end, one again obtains \er{integral12}.

\subsection {The Yang-Mills fields and collected results}

These are defined in \er{YMF1}, are shown for the adiabatic
representation
in \er{F11} - \er{F12}) and have the following forms:
\ber
{\bf F}^b_{11}&=& (bZ/2qR_b)\delta (q) {\bf\hat Z}\\
&[=&(\pi (bZ/R_b)\delta (X)\delta (Y)){\bf k}] \\
&=&- {\bf F}^b_{22}
\label {F11ad}
\enr
and
\ber
{\bf F}^b_{12}&=&-{1\over2} \frac{\delta (q)}{R_b}{\bf\hat Z} \\
&[=&-\pi q/R_b)\delta (X)\delta (Y) {\bf k}]
\nonumber\\
&=& {\bf F}^b_{21}
\label {F121ad}
\enr
These are much simpler than the magnetic (purely curl) field
expressions.
Their fluxes are given below, in Table 1, where we collect all results
in
the 2-parameter  $b \to 0$ limit.

-----------------------------------------------------------------------
--------

Table 1.Summary for conical intersections in the adiabatic
representation in
 the two-dimensional, $b \to 0$ limit (designated by 0 sub- or super-
script.)

Derivatives:${\bf\vec \nabla}\phi ={\bf\hat\phi}/q,{\bf \vec
\nabla}\theta_0 =\pi
 \delta(q) {\bf\hat q},{\bf\vec \nabla}\theta_0 \wedge {\bf\vec
\nabla}\phi=
\pi \delta (q)({\bf\hat q}/Z +{\bf\hat Z}/q)$

NACT: ${\bf\tau}^0_{11}=-{\bf\tau}^0_{22}=-(i/2)\pi
 \delta(q) {\bf\hat \phi}; {\bf\tau}^0_{12}={\bf\tau}^{0*}_{21}=
{i\over2}{\bf\hat\phi}/q - {1\over2}\delta(q) {\bf\hat q}$

Flux or line integrals:

$(Magnetic flux)_{11}=-(Magnetic flux)_{22}=0$

$(Magnetic flux)_{12}=(Magnetic flux)_{21}=-\pi$

$(YM flux)11=- (YM flux)22 = \pi$

$(YM flux)12= (YM flux)21= 0$

-----------------------------------------------------------------------
--------

As already noted, the most remarkable results in Table 1 are the non-
zero
values of the NACTs upon the seam $q=0$ and the circumstance that the
magnetic
and Yang-Mills fluxes appear in an opposite manner. These are the
legacy of our
using the $b \to 0$ limiting procedure, rather than starting with a
Hamiltonian
in which $b=0$, the $Z$-coordinate is absent and the seam $q=0$ is a
line of
 singularity.
 The  observational significance of the fields or fluxes will be
discussed later.

\subsection {Circulating representation}

In the foregoing, we have calculated the field and the flux on the
assumption
 that the electron is in an adiabatic state. For a slowly changing
perturbation,
such a state is  a
 stationary state of the system.The diagonal elements of the tensorial
 quantities that we have calculated refer to this situation. In
principle, the
 electron may be excited to a state different from an adiabatic one,
e.g., to
 a linear combination of two adiabatic states with constant
coefficients. The
 superposition of adiabatic states will persist, since under
  quasi-stationary
 conditions each adiabatic state will develop in an independent
fashion. The
 fields pertaining to such a situation can be obtained through use of
the
 non-diagonal elements of the field (or flux), calculated in this work.
 However, the magnetic field can also react on the electronic motion
and this reaction
 might cause changes in the adiabatic state. This effect is formulated
in the
 last subsection.

 We now calculate the fields in a different representation, which we
name
 "circulating representation" and denote by circularly shaped bras and
kets,
 $|+), |-)$. This is the representation introduced by Baer \cite
{Baer2001} so
as to obtain NACTs that are diagonal (pseudoscalar) and purely
imaginary. (It
is suggested that in a non-reactive scattering, the  $|+), |-)$ states
 resemble the  circulating-state situation.)

The following transformation, acting on the adiabatic states in
\er{1adstate}, \er{2adstate}, generates the circulating representation:
\beq
|+)= [|1>+|2>]/\sqrt 2
\label {2circstate}
\enq
\beq
|-) = [|1>-|2>]/\sqrt 2
\label{1circstate}
\enq
In the limit $b \to 0$ (where
$\sin{\frac{\theta_0}{2}}=\cos{\frac{\theta_0}{2}}
 =1/\sqrt 2$), one has
\ber
|+)= e^{i\frac{\phi}{2}} |-> \\
|-)= e^{-i\frac{\phi}{2}} |+>
\label{limitcircular}
\enr
where, we recall, $|->$ and $|+>$ are diabatic coordinate independent
 electronic  states. The complex
 exponential prefactor is the reason for the name "circulating
 representation". [When $b$ is non-zero, the circulating states are
coordinate
dependent superpositions of the diabatic states.]
We list the results in this representation, obtained after using some algebra:
\vspace{2cm}

-----------------------------------------------------------------------
--------

Table 2. Results in the circulating representation with the arrow
denoting
the $b \to 0 $ limit:

NACTs:

 ${\bf\tau}^b_{++}= \frac{i{\bf\hat\phi}}{2R_b} \to
\frac{i{\bf\hat\phi}}{2q}$

${\bf\tau}^b_{--}= -\frac{i{\bf\hat\phi}}{2R_b} $

${\bf\tau}^b_{-+}= -\frac{i{bZ{\bf\hat\phi}}}{2qR_b} -
{1\over2}\frac{bZ}{R_b^2}{\bf\hat q}
+{1\over2}\frac{b q{\bf\hat Z}}{R_b^2}$

${\bf\tau}^b_{+-}= -\frac{i{bZ{\bf\hat\phi}}}{2qR_b} +
{1\over2}\frac{bZ}{R_b^2}{\bf\hat q}
-{1\over2}\frac{b q{\bf\hat Z}}{R_b^2}$

Fluxes or line integrals, for $b\to 0$:

$(Magnetic flux)_{++}=- (Magnetic flux)_{--}\to -\pi$

$(Magnetic flux)_{+-}=(Magnetic flux)_{-+} \to 0$

$(YM flux)_{++}=-(YM flux)_{--} \to 0 $

$(YM flux)_{+-}= (YM flux)_{-+} \to \pi$

-----------------------------------------------------------------------
--------

We note that for $b$ nonzero the NACTs have also off-diagonal elements
and that
on the seam $q=0$ these elements exist even in the two-dimensional
limit $b \to
0$. Moreover, these elements have components other than tangential
 (along $ {\bf\hat\phi})$.
 As such, they do not contribute to the line integral in \er
{integral}.
At finite distances from the seam $q>0$, our results agree with those
in
\cite {Baer2001}.

\subsection{An alternative formalism}
    Our results could also have been obtained, had we proceeded
 differently, namely by making the $X$ and $Z$ the "active" variables
and $Y$
 the fictitious one. To achieve this one puts the $b$ factor (whose
limit 0
 is  ultimately taken) in front of $Y$ in the Berry Hamitonian, our \er
{BerryH}, rather than before $Z$. This amounts to a representation of
the
diabatic electronic
 states that is obtained from the $|+>, |->$ set used above, by a
complex
 transformation. In the $b\to 0$ limit the adiabatic states are the
same in
 either formalism
  and so are the NACTs and the magnetic field. An historical interest
is
 attached to the latter procedure in that Stone not only proposed this
 Hamiltonian \cite {Stone}, with the interpretation of $b$ as a spin
orbit
 coupling  strength, but even considered the limit of $b\to 0$ .
However, he
 did not derive a magnetic field.
\subsection {General linear coupling and derived quantities}
The most general form for the interaction Hamiltonian that is linear in
the
parameters $(X,Y,Z)$ is the following:
\beq
H_b  = \left(\begin{array}{cc}
           bZ & \alpha X-i\beta Y\\
           \alpha X+i\beta Y & -bZ
\end{array} \right)\qquad\\
\label {ellipticH}
\enq
This differs from \er{BerryH} by the inequivalence between not only
$Z$ (for $b \ne 1$)
 and $X,Y$, but also between $X$ and $Y$. One regains the formalism
of \cite{BaerME}, upon using the substitutions $b=0, \alpha \to b,
\beta \to 1$. In
this two-dimensional (or two-parameter) case the intersection between
the two
adiabatic potential surfaces (the solution of \er{ellipticH}) involves
two
inverted elliptic cones, rather than  two inverted circular ones which
one
obtains for $b=0, \alpha=\beta$.
The method of the previous section can be easily carried over to this
case
and we shall only quote the modifications needed for the
generalization. Our main
purpose is to demonstrate explicitly the dependence of the  magnetic
and Yang-Mills fields
on the angle $\phi$ defined in \er{phi}. A dependence of this type
 has been  predicted in \cite {Baer2001}.
We first define
\beq
\gamma=\alpha / \beta
\label{gamma}
\enq
so that $\gamma= 1, \beta=1$ takes us back to the previous sections.
Otherwise,
(when $\gamma \ne 1$)we have the following modifications, designated by
placing
an apostrophe over all symbols affected:

The adiabatic states have now the following form:
\beq
|1'> =  e^{-i\phi' /2} \cos(\theta '_b /2)|+> + e^{i\phi' /2}
\sin(\theta '_b /2)|->\\
\label {1adstatea}
\enq
\beq
|2'> = -e^{-i\phi' /2} \sin (\theta '_b /2) |+> + e^{i\phi' /2}
\cos(\theta '_b /2)|->
\label {2adstatea}
\enq
where
\beq
R'_b =\sqrt{(\alpha X)^2 +(\beta Y)^2+(bZ)^2} =\sqrt {q'^2+(bZ)^2}
\label {Rab}
\enq
\beq
tan\phi '  = \beta Y/(\alpha X)
\label {phia}
\enq
\beq
tan \theta '_b =q'/bZ
\label {thetaa}
\enq
having introduced  the modified radial cylindrical coordinate $q'$
through
\beq
q'^2=(\alpha X)^2 +(\beta Y)^2
\label {qa}
\enq

The NACTs  are now
\ber
\bf\tau ^{'b}_{11} & \equiv & <1'|\vec \nabla|1'> = -{i\over
2}\cos\theta ' _b
\vec \nabla\phi ' \\
                & =&-\bf\tau  ^{'b}_{22}
\label{tau11a}
\enr
\beq
{\bf\tau}  ^{'b}_{12}  = {1\over 2}i \sin\theta' _b \vec \nabla\phi ' -
{1\over2}\vec \nabla\theta' _b
\label {tau12a}
\enq
\beq
{\bf\tau}  ^{'b}_{21}  = {1\over 2}i \sin\theta' _b \vec \nabla\phi '+
{1\over2}\vec \nabla\theta' _b
\label {tau21a}
\enq
The derivatives of the angles are
\beq
{\bf\vec \nabla}\phi ' = \frac{\gamma}{1+(\gamma ^2 -1)\cos^2\phi}{\bf
\hat\phi} /q
= \frac{\alpha \beta}{q^{'2}} q {\bf \hat\phi}
\label{nablaphia}
\enq
\beq
{\bf\vec \nabla}{ \bf\wedge\vec \nabla}\phi ' = \frac{\gamma}{1+(\gamma
^2 -1)\cos^2\phi}\frac{\delta(q)}{q} {\bf{\hat Z}}
=\frac{\alpha \beta}{q^{'2}}q \delta(q) {\bf\hat Z}
\label {curldivphia}
\enq
\beq
{\bf\vec \nabla}\theta '_b = -(bq'/R_b^{'2}){\bf\hat Z}
+(bZ/R_b^{'2})\frac{q'}{q}
{\bf\hat q} -\frac{bZq}{2q' R_b ^{'2}}(\alpha ^2 -\beta ^2) \sin 2\phi
{\bf\hat \phi}
\label {nablathetaa}
\enq
and
\beq
{\bf\vec \nabla }\theta '_b \wedge {\bf\vec \nabla }\phi '
= \frac{bZ}{R_b^{'2}} \frac{\alpha\beta q}{q'} ({\bf \hat q} /Z +{\bf
\hat Z} /q)
\label{nablanablaa}
\enq
In \er{nablathetaa} the last term is new. In the adiabatic
representation
\er {1adstatea} and \er {2adstatea} one finds the following NACTs
\ber
{\bf\tau} ^{'b}_{11} & =& -{i\over 2}
(bZ/qR'_b)\frac{\gamma}{1+(\gamma^2-1)
\cos^2 \phi}{\bf\hat\phi}\\
& =&-{\bf\tau} ^{'b}_{22}
\label{tau11ada}
\enr
and
\ber
{\bf \tau}^{'b}_{12} & =& \big( \frac{i\alpha \beta}{2R'_b}
+(\alpha^2-\beta^2)\sin2\phi \frac{bZ}{4R^{'2}_b}\big) \frac{q}{q'}
{\bf\hat\phi}
 -\frac{bZq'}{2R_b^{'2}}(\frac{{\bf\hat q}}{q}
 -\frac{{\bf\hat Z}}{Z})\\
{\bf \tau} ^{'b}_{21} & =& \big( \frac{i\alpha \beta}{2R'_b}
-(\alpha^2-\beta^2)\sin2\phi \frac{bZ}{4R^{'2}_b}\big) \frac{q}{q'}
{\bf\hat\phi}
 +\frac{bZq'}{2R_b^{'2}}(\frac{{\bf\hat q}}{q}
 -\frac{{\bf\hat Z}}{Z})
\label {tau12ada}
\enr
In the two parameter limit $ b \to 0$ for non-zero $q$ the only term
 remaining is
\ber
{\bf \tau} ^{'0}_{12}&=&  {\bf \tau} ^{'0}_{21}\\
 & =&  \frac{i\alpha \beta}{2R'_b}
 \frac{q}{q'} {\bf\hat\phi}\\
&=& {i\over 2} \frac{\gamma}{1+(\gamma^2-1)\cos^2
\phi}\frac{{\bf\hat\phi}}{q}
\label {tau12ada0}
\enr
The off-diagonal coupling term obtained in \cite {MebelBL2000}
computationally
for the molecule $C_2H$ near
 a conical intersection was subsequently fitted to the above expression
in
 \er{tau12ada0}
 \cite {BaerME}.  The coupling term is characterized (for $\gamma
<1)$by two (frequently quite dominant) peaks
at $\phi=0(= 2\pi)$ and $\pi$. However, on the seam $q=0$ there are
additional
 non-zero
terms in  ${\bf\tau} ^{'0}_{12}$, in ${\bf\tau} ^{'0}_{11}$, etc., as
we have
 already noted. These are new results. One expects similar angular
behavior from them, too.
The line integral over the coupling coefficient,  or the  Berry phase,
in
 \er{integral} gives in the $b \to 0$ limit the values of 0 for the
diagonal
terms and $\pm \pi$ for the off diagonal terms, irrespective of the
value
of the ratio $\gamma =\alpha/\beta$. These values were confirmed
numerically to a good
 approximation for  the neighborhood of the intersection in \cite
{MebelBL2000} and discussed in \cite {BaerME}.

Returning to  general values of the parameters $b,\alpha, \beta,$ we
show now
 the fields (magnetic and Yang-Mills):
\ber
{\bf H}^{'b}_{11}&=& \alpha \beta
[\frac{bZ}{2R_b^{'2}}(\frac{qR'_b}{q^{'2}}
\delta (q) - \frac{1}{R'_b}){\bf\hat Z} -
 (bq/2R_b^{'3}){\bf\hat q}]\\
&=&- {\bf H}^{'b}_{22}
\label {H11ada}
\enr
\ber
{\bf H}^{'b}_{12}&=&-\alpha \beta [( \frac{q\delta (q)}{2q'R'_b}
+\frac{(bZ)^2}{2q'R_b^{'3}}){\bf\hat Z} -
\frac{b^2 Z}{2R_b^{'3}}\frac{q}{q'} {\bf\hat q}]
\label{H121adb}\\
&=& {\bf H}^{'b}_{21}
\label {H121ada}
\enr

In the limit $b \to 0$, this field also vanishes outside $q=0$. However,
on the seam
$q=0$ we obtain one of the interesting results of  this section, namely
the angular
dependence of the magnetic field. Only the second  term in \er
{H121adb} for ${\bf H}^{'b}_{12}$
survives in this limit and one obtains after a slight simplification
\beq
{\bf H}^{'0}_{12}= -{1\over2} \frac{\gamma \delta (q)}{1+(\gamma^2-
1)\cos^2 \phi }{\bf\hat Z}
\label{H120ada}
\enq
An angular dependence of the seam magnetic field was postulated in
\cite{Baer2001}.
The above relation gives its form within the general (elliptic)
linearly dependent model.
 The total flux in the $b \to 0$ limit is again the limiting form of
 \er{integral12}, namely $-\pi$.

The Yang Mills tensorial  fields are calculated as:
\ber
{\bf F}^{'b}_{11}&=& \alpha \beta \frac{bZq\delta (q)}{2q^{'2}R'_b }
{\bf\hat Z}\\
&=&- {\bf F}^{'b}_{22}
\label {F11ada}
\enr
whose integrated flux in the $b \to 0$ limit is $\pi$ for all values of
 $\alpha$ and $ \beta $ . Finally,
\ber
{\bf F}^{'b}_{12}&=&- \alpha \beta \frac{q\delta (q)}{2q'R_b}{\bf\hat
Z} \\
&=& {\bf F}^{'b}_{21}
\label {F121ada}
\enr
whose flux is 0 for all finite values of $b$. Since in our limiting
procedure
the limit $b \to 0$ is taken at the end, we have a zero flux in the
two-dimensional
elliptic geometry, which is identical to the entry in Table 1 for the circular
case.

\subsection {Interpretation of the fields}
Associated with a {\it ci} of two potential surfaces for a
 polyatomic molecule, there exists an analogue of a magnetic field,
which
 affects the nuclear motion through its presence in the nuclear
Schrodinger
equation.  However, in addition to the magnetic field,
there is  an analogous, symmetry-based field, the Yang-Mills or
tensorial
 field. By  employing a limiting procedure we have shown that for a
(multiple valued) adiabatic state both types of the field have
delta-function-like, thin  solenoidal forms.

 In the molecular context, the direction of the solenoid is defined by
the
 {\it ci}, as follows: The {\it ci} defines a
plane in the nuclear
 coordinate space; the direction of the field is along any arbitrary
 direction in the nuclear coordinate space (the "seam"), which is
 perpendicular to this plane. Though the two types of fields have
similarities,
 they differ in their numerical values
 and the fluxes due to them "complement" each other (Tables 1 and 2).
We have also found that the fluxes are quantized, meaning that the
strength
 of the field and the flux associate with it do not depend on physical
 parameters, just as they don't for the magnetic monopole field of
Dirac
 \cite{Dirac31}.
 This result was obtained recently \cite{Baer2001}, where the "curl-
field"
 was calculated by applying a complex-valued linear superposition of
the
 adiabatic states. This superposition is equivalent to that introduced
here
 for a pair of
 circulating states (one clockwise and another anti-clockwise) around
the
seam. It also expresses the geometry of the Aharonov-Bohm effect, in
which
 two currents circle around a screened solenoid in opposite senses.
However, in
 that case  the flux is not quantized, but depends on the magnetic
field inside
the solenoid.

 As noted above,
 the "curl-field" is a delta function along the seam, so that a
particle circulating at a finite radius
 would  be oblivious of this field.
A different situation could arise when the electron is excited into a
general
 superposition state, i.e. one that is not a superposition of adiabatic
 states (or {\bf is} a superposition thereof, but with coefficients
depending on the
 coordinates.) The "magnetic field" would be completely different from
those
 for the adiabatic states.  Formally, the new, general superposition
would be
 described by applying what is called a "non-local" gauge
transformation
 \cite {Jackiw, Weinberg}.
 The effect of this is well known and is expressed by saying that the
vector
 potentials $\bf A$ (or the NACTs) transform inhomogeneously and the
tensorial field
 $\bf F$ does so homogeneously in a covariant way \cite {Jackiw}.
However,
 for a coordinate dependent superposition, the fields may be difficult
observe
 (if at all possible), due  to the off-diagonal matrix elements in the
potential, which make such a  state non-stationary.
\subsection {Observational aspects through effective Hamiltonians }
\label{Observational aspects through effective Hamiltonians }
Possible experimental consequences of gauge fields have been noted for
 electron spin experiments with time-varying magnetic fields
\cite{MoodySW},
in atoms with rotating electric fields \cite{MoodySW}, in collisions
between
 atoms \cite {Zygelman} and in further applications \cite {ShapereW}.

We now give a general formalism for the observational effects of the
fields,
 one that holds the promise of differentiating between the magnetic and
the
 Yang-Mills field. It is based on an effective or truncated Hamiltonian
 formalism, similar in many respects to the well known
 Spin-Hamiltonian description (\cite {Pryce} -\cite {Stoneham}). This
concentrates
on a small set of states (in the present context, the two-fold  set in
\er{1adstate}
 and \er {2adstate}), and considers the effects of perturbations on
these.
 The perturbations admix states from outside the small set.
 The Spin-Hamiltonian formalism  shows a way to represent
the effect of the full set within the small ("truncated") set, in such
a manner that
the excited electronic states are included only "virtually". Symmetry
considerations
determine the form of the truncated Hamiltonian. Clearly, since several
independent expressions can be compatible with the symmetry
requirement, as e.g., by
including higher order effects, the effective Hamiltonian  will
normally consist of
several terms. The coefficients with which these terms enter will in
general
not be amenable to calculations, but are empirically determined.

 After these preliminary remarks we recall that the previous
subsections
 treated electronic
and nuclear degrees of freedom. The fields that we have derived were
pseudo-vector
quantities in the nuclear space and were functions of the nuclear
variables
$(X,Y,Z)$ or $(q, \theta, \phi)$. Moreover,
the fields were elements of a matrix (or tensor) in the Hilbert space
of the
electronic set, which is not the case in ordinary electromagnetism. We
shall
 write a generic field component (not differentiating for the
moment between magnetic and  Yang-Mills fields) as
\beq
F^a_{mn}=F^a_{mn}(X,Y,Z)
\label{field}
\enq
where $a$ is the vector-index in the nuclear coordinate space and
$(m,n)$ are
 labels for the electronic set. We shall also consider operators of two
types.
First, the operators
\beq
op^a_1, op^{ab}_2, etc.
\label{operator}
\enq
in the nuclear vector-space that are functions of electronic variables.
It is now elementary to construct terms for an effective Hamiltonian
$H_{eff}$,
 such that satisfy the  symmetry requirements:
\beq
H_{eff}=  |m>(C_1 F^a_{mr}<r|op^a_1|n> + C_2 F^a_{mp}F^b_
{pr} <r|op^{ab}_2|n> + ...)<n|
\label{effH}
\enq
with possible additional terms to follow. Summation for repeated
indexes is
 implied. The
first operator $op^a_1$ is a pseudo-scalar and the second operator
$op^{ab}_2 $
has also the appropriate transformation properties.(Thus it might be a
direct
product of two $op^a_1$'s.) The coefficients $C_1, C_2, etc. $ are
expected to be
empirical parameters.

As a second application we consider the extension of the electronic
(orbital) degrees
of freedom, e.g., by inclusion of electronic spins. We label the set
for the
extended degree of freedom by $|M>$ and  operators that act on both the
electronic orbital and spin degrees of freedom by $Op^a_1, Op^{ab}_2$,
etc.
The $a,b$ indexes refer to the axes in nuclear space, as before. Then
the effective
Hamiltonian takes the following form:
\ber
H_{eff}=  |m>|M>(C_1 F^a_{mr}<r|<M|Op^a_1|n>|N>
\nonumber\\
+ C_2 F^a_{mp}F^b_{pr} <r|<M|Op^{ab}_2|n>|N> + ...)<n|<N|
\label{effH2}
\enr
Evaluation of these Hamiltonians requires computing expectation values
in
a given nuclear state. Since $H_{eff}$ is anticipated to be small
compared to
energies of the nuclear freedom, this is a legitimate procedure.
The effective Hamiltonian thus provides a way to include residual
perturbations
that couple to states outside the degenerate electronic doublet. We
have seen
above (e.g. in Tables 1 and 2) that the magnetic field and the Yang-
Mills
field differ markedly. It is therefore suggested that by experimentally
testing the effective Hamiltonian (e.g., through its dependence on the
nuclear
vibrational levels) one could establish which field is effective.
\section {Conclusion}
This paper explains the signs of topological
phases obtained by various authors in several previous works. The sign
 depends on the
{\it derivatives} of the coupling matrix and is  shown by the
inequalities in \er {xineq} for
a cartesian, real representation and in \er {theta31} for a complex
representation.
Although positively and negatively signed topological phases cannot be
distinguished, since they have the same phase factor, this is true only
upon
performing a complete loop. For loops shorter or longer that this, the
sign
of the phase change is observationally accessible.

We then obtain for various models the state- or representation-
dependent magnetic and Yang-Mills fields, which result from the Born - Oppenheimer scheme for
the coupling between nuclear and electronic degrees of freedom. An
effective or truncated Hamiltonian, here suggested, provides a possible tool for
experimental verification of the fields.

\section {Figure Caption}
Figure 1.

Coordinate systems for the point $ (X,Y) $ or $ (q,\phi ) $ that
circles around
the point of conical intersection [the dot located at $ (X_0,Y_0) $ or
$
(q_0,\phi_0 )$ ].
The circling is with a radius $ \Delta $, and circling angle $ \alpha
$.

\end{document}